\documentclass[fleqn,10pt]{wlscirep}
\usepackage[utf8]{inputenc}
\usepackage[T1]{fontenc}
\usepackage{multirow}
\usepackage{soul}
\usepackage{cancel}
\def\track#1{{\color{black}#1}}
\title{Social norms in indirect reciprocity with ternary reputations}

\author[1]{Yohsuke Murase}
\author[2]{Minjae Kim}
\author[2,*]{Seung Ki Baek}
\affil[1]{RIKEN Center for Computational Science, Kobe, Hyogo 650-0047, Japan}
\affil[2]{Department of Physics, Pukyong National University, Busan 48513, Korea}

\affil[*]{seungki@pknu.ac.kr}

\begin{abstract}
Indirect reciprocity is a key mechanism that promotes cooperation in social
dilemmas by means of reputation. Although it has been a common practice to
represent reputations by binary values, either `good' or `bad', such a dichotomy
is a crude approximation considering the complexity of reality. In this work, we
studied norms with three different reputations, i.e., `good', `neutral', and
`bad'. Through massive supercomputing for handling more than thirty billion
possibilities, we fully identified which norms achieve cooperation and possess
evolutionary stability against behavioural mutants. By systematically
categorizing all these norms according to their behaviours, we found
similarities and dissimilarities to their binary-reputation counterpart, the
leading eight.
We obtained four rules that should be satisfied by the successful norms, and the
behaviour of the leading eight can be understood as a special case of these
rules. A couple of norms that show counter-intuitive behaviours are also
presented. We believe the findings are also useful for designing successful
norms with more general reputation systems.
\end{abstract}
\begin{document}

\flushbottom
\maketitle
\thispagestyle{empty}

\section*{Introduction}

The ability to cooperate with others that are genetically unrelated is a remarkable trait of humans.
Reputation is formed by evaluating each other after observing who does what against whom, which in turn incentivizes an action that is costly but beneficial to others because those with good reputations are likely to receive benefits subsequently in society.
This is known as indirect reciprocity, one of the most fundamental mechanisms for maintaining cooperation~\cite{nowak2006five}.
Whether a given action is perceived as good depends on the action itself, the context, and the social norm used by the observer.
A central question is thus what are the requirements of social norms to
achieve social cooperation.

The leading eight is a set of highly successful social norms for maintaining
cooperation at a high level~\cite{ohtsuki2004should}.
By comprehensive enumeration of possible social norms, it has been found that the
leading eight are the only ones that can sustain evolutionarily stable
cooperation for a broad range of the benefit-to-cost ratios of cooperation.
Because of its simplicity and effectiveness, the leading eight have served as a
baseline in a wide range of theoretical studies of indirect
reciprocity~\cite{ohtsuki2009indirect,uchida2013effect,martinez2013evolutionary,santos2016social,santos2018social,okada2017tolerant,okada2018solution,hilbe2018indirect}.

Most of the previous studies on indirect reciprocity, including the leading
eight, assume that the reputation of a player is represented by either
`good'($G$) or `bad'($B$)~\cite{nowak2005evolution,ohtsuki2006leading,suzuki2007evolution,uchida2010competition}.
Whereas the assumption of binary reputation has been widely adopted as a common practice
for its simplicity and theoretical tractability,
such dichotomy is not always realistic given experimental evidence and our daily experience.
Furthermore, it is not always clear how much we can generalize the conclusions
obtained from the binary-reputation models to more
realistic and complex reputation models because the conclusions may be
consequences of the oversimplification.

What would be the universal characteristics that every successful norm shares
irrespective of the form of reputation?
How should we revise the conclusion learned from the
binary-reputation system when the binarity assumption is relaxed?
Answering these questions has been a serious challenge because
the strategy space expands super-exponentially
with the number of possible values of reputations:
The number of third-order social norms with $k$ reputations is about $(2k^2)^{k^2}/k!$ \track{as we will discuss in the next section}.
Although there are several studies that go beyond the binary assumption,
only a small subset of the norms were studied in these studies by assuming ordinal relationships between reputations~\cite{tanabe2013indirect,nowak1998evolution,nakamura2011indirect,berger2011learning,berger2016stability,clark2020indirect,schmid2021unified,lee2021local}.
In particular, a continuum formulation of indirect reciprocity allows a perturbative analysis,
from which one can derive a condition for linear stability against erroneous disagreement,
but it is applicable only to mutants that are sufficiently close to the resident norm~\cite{lee2021local}.

In this study, to bridge the gap between the binary and more general models of reputation, we comprehensively study the norms with a ternary-reputation model \track{under public reputation},
in which players are labelled by three types of reputations.
The ternary counterparts of the leading eight will be fully identified by
comprehensive enumeration of the third-order social norms
through state-of-the-art supercomputing.
Such a large-scale enumerative approach has also proved useful in studies of direct
reciprocity~\cite{yi2017combination,murase2018seven,murase2020five},
and this study is an application of the method to the study of indirect reciprocity.
As we will see in the following, both similarity and dissimilarity are found
between binary- and the ternary-reputation models, indicating potentially
universal features in indirect reciprocity as well as limitations of the binary
reputation.
We also obtained some counter-intuitive results that would not have been discovered
without computational methods.

This paper is organized as follows: In the next section, we present the model
definition and an outline of the algorithm to find successful social norms.
In the result section, the resulting ternary norms are presented after
we briefly review the characteristics of the leading eight.
To better comprehend the working mechanisms, we classify the norms
by observing their differences.
Some counter-intuitive examples are also presented.
Finally, in the last section, we discuss the similarities and dissimilarities with
respect to the leading eight and summarize this paper.

\section*{Model}

\subsection*{Description}

In this study, we closely follow the previous settings for the leading eight~\cite{ohtsuki2004should}
but with ternary reputations.
We consider an infinitely large population. In each round, two players are
randomly
picked to form a donor-recipient pair and play the one-shot
donation game. The donor decides to either cooperate ($C$) or defect ($D$)
considering the recipient's social reputation as well as his or her own.
Cooperation costs the donor a payoff loss $c>0$, whereas the recipient receives a
benefit $b (>c)$. On the other hand, defection means that the donor does nothing
to the recipient. The donor is always better off by choosing $D$. Hence
the game represents a social dilemma.
This process is repeated sufficiently long to reach a stationary state, until which
players have engaged in the donation game many times with different opponents.
%\track{\st{Such a stationary state may also be called an equilibrium.}}

We assume public reputation, by which we mean that everyone in a population
shares the same assignment rule and thus assigns the same reputation to each
player.
\track{In other words, any reputation assigned to an individual is shared by
all members of the population in complete agreement.}
We have three labels for representing reputations, i.e., $G$ (good), $N$
(neutral), and $B$ (bad). However, note that we do not assume any {\it a priori} ordinal
relationships among them so that $N$ may be interpreted as worse than $B$, for
instance.

A social norm is comprised of an assessment rule and an action rule,
and both the rules are assumed to be
deterministic: Concerning the assignment part, it means that
a donor's new reputation $Z \in \{G,N,B\}$ is a function of the donor's
reputation, the recipient's reputation, and the donor's action. In other words,
an assignment rule is represented by a map $R(X,Y,A) \to Z$, where $X,Y \in
\{G,N,B\}$ are reputations of the donor and the recipient, respectively, and $A
\in \{C,D\}$ is the donor's action. Likewise, an action rule is represented by a
map $P(X,Y)\to A$, where $X$ and $Y$ are reputations of the donor and the
recipient, respectively, and $A$ is the prescribed action. In the following, the
combination of assignment and action rules is denoted as a norm.

\track{When $k$ reputations are available, the number of norms is approximately
$(2k^2)^{k^2}/k!$:
To define an assignment rule $R(X,Y,A) \to Z$, $Z$ must be determined out of $k$
possibilities for each combination of $(X,Y,A)$.
Thus the number of assignment rules is $k^{2k^2}$.
Similarly, to define an action rule $P(X,Y) \to A$, one has to determine $A \in
\{C,D\}$ for each combination of $(X,Y)$, yielding $2^{k^2}$ action rules.
Therefore, the total number of social norms is $(2k^2)^{k^2}$.
Note that we have counted all the norms even if some of them are equivalent with
respect to the permutation of $k$ types of reputation.
For example, if we swap $G$ and $B$ everywhere, all essential
predictions derived from the model will remain unchanged because reputations are
mere labels to distinguish $k$ social states.
Taking this permutation into account, the number of social norms is described by
$(2k^2)^{k^2}/k!$. However, this is a rough estimate
because the norms that are symmetric under permutation would not be
counted $k!$ times~\cite{ohtsuki2004should}.}

Both implementation error and assignment error are included in our
calculation:
With probability $\mu_e$, a player defects when the prescribed action is
cooperation. The opposite error does not occur, i.e., a player does not
cooperate by mistake when the prescribed action is defection.
An assignment error occurs with probability $\mu_a$. When an assignment error
occurs, the donor is assigned a different reputation from the prescribed
one. For instance, either $B$ or $N$ will be assigned to a donor with equal
probability ($=\mu_a/2$) when the assignment rule prescribes $G$.
Following the indirect observation model~\cite{ohtsuki2004should},
a misperception will also be shared by the population.

\subsection*{Calculation}

We will identify all the norms that form strict Nash equilibria with sufficiently high cooperation rates.
Because a strict Nash equilibrium implies an evolutionarily stable strategy
(ESS)~\cite{smith1982evolution}, we call each of them
a \emph{cooperative ESS} or a \emph{CESS} for short.
Because of $18$ possible combination of $(X,Y,A)$, we have $3^{18} =
387,420,489$ different assignment rules in total.
Taking the permutation symmetry of reputations into
account~\cite{ohtsuki2004should}, \track{we computationally found that
the number of independent assignment rules reduces to $64,573,605$.}
For each assignment rule, $2^{9} = 512$ possible action rules exist.
Thus, the number of independent pairs of assignment and action rules amounts to
$33,061,685,760$.
We judge whether a norm $S$ is a CESS by using the
following algorithm (see Methods for more details):
\begin{enumerate}
    \item Let $h_B$, $h_N$, and $h_G$ denote respective fractions of $B$, $N$,
    and $G$. Calculate their values in \track{a stationary state}, denoted by $h_B^\ast$,
    $h_N^\ast$, and $h_G^\ast$, respectively, under the assumption that the
    entire population uses $S$.
    \item Calculate the cooperation level $p_{c}$, which means the probability
    that a donor cooperates towards a recipient when both are randomly picked
    from the resident population.
    \item Reject the norm if $p_c < p_c^{\rm th}$, where $p_c^{\rm th}$ is a threshold for the cooperation level.
    \item Otherwise, calculate the payoff of a mutant with a different action
    rule from the resident one under the assumption that mutants occupy a
    sufficiently small fraction.
    \item Repeat the above step for all possible action rules. If the payoff of
    a resident is higher than that of any possible mutants, $S$ is a CESS.
\end{enumerate}
%The codes and the results for this study are available
%online\footnote{\url{https://github.com/yohm/sim_game_ternary_reputation}}.

Here, we define a CESS as a norm whose defection level $p_d \equiv 1-p_c$ scales as $O(\mu_a)+O(\mu_e)$ as $\mu \to 0$,
that is, the probability of prescribing defection is of the same order as the error rates.
Some norms show slower convergence, such as $p_d = O(\mu_a^{1/2})$,
both in the binary- and the ternary-reputation cases, and we do not include such norms that are fragile against noise.
This requirement is consistent with the criteria for finding the leading eight in the binary-reputation case.
In the following calculation, we use $\mu = 10^{-3}$, where $\mu \equiv \mu_e = \mu_a$, and $p_c^{\rm th} = 0.99$.
We choose these values such that $1-p_c^{\rm th}$ is sufficiently larger than $\mu$ but smaller than $\sqrt{\mu}$.
For the CESS's found in the following, we numerically confirmed that $p_d = O(\mu)$ by calculating the cooperation levels for different values of $\mu$.

% Recall that we assume no ordinal relation among $G$, $N$, and $B$. For this
% reason, the list of CESS's should include equivalent norms up to their
% permutation.
% To remove such trivial multiplicity, we relabel the reputation with
% the largest fraction as $G$.
As mentioned above, the norms generated by permuting $G$, $N$, and $B$ are equivalent because we assume no ordinal relations among them.
To remove trivial multiplicity, we use the following protocol:
First, the reputation with the largest fraction is labelled as $G$.
In other words, we always have $h_G^{\ast} > h_N^{\ast}$ and $h_G^{\ast} > h_B^{\ast}$.
To be a CESS, therefore, the action
rule must achieve high $p_c$ by prescribing $C$ when both the players have
reputation $G$. If this rule is violated by mistake, the donor must get a
reputation other than $G$ because, otherwise, the donor would not find any incentive to
cooperate.
We define the reputation resulting from such defection as $B$, and
the last remaining one as $N$.
The labels assigned by these guidelines are overall consistent with our common
sense of `good', `neutral', and `bad', as we will see in the following.

% Results and Discussion can be combined.
\section*{Results}

\subsection*{Leading eight in the binary-reputation model}

Before showing our results for the ternary-reputation model, let us review the
characteristics of the leading eight. We will check whether these are
universally shared with successful norms in the ternary-reputation model.
By the leading eight, we mean eight norms that qualify as CESS's
in the binary-reputation model, and they are characterized by the
following four properties~\cite{ohtsuki2006leading}:
\begin{enumerate}
    \item Maintenance of cooperation: $P(G,G) = C$ and $R(G,G,C) = G$.
    \item Identification of defectors: $R(\ast,G,D) = B$.
    \item Punishment and justification of punishment: $P(G,B) = D$ and $R(G,B,D)
    = G$.
    \item Apology and forgiveness: $P(B,G)=C$ and $R(B,G,C) = G$.
\end{enumerate}
With the leading eight, the community is mostly occupied by a single type of players
($G$) who form mutual cooperation, whereas the fraction of $B$-players is of $O(\mu)$.
When someone defected from cooperation, the population assigns
reputation $B$ to the defector to distinguish him or her from cooperators.
$G$-players punish such a $B$-player by refusing cooperation, and the punishment
is justified in the sense that the defection does not hurt their $G$-reputation.
A $B$-player can obtain $G$-reputation by donating to a $G$-player as an
apology. The prescriptions of the leading eight are summarized in
Table~\ref{tab:leading_eight}.
\begin{table}[ht]
  \centering
  \caption{
  Prescriptions that are commonly shared by the leading eight. The asterisk
  ($\ast$) is a wildcard, meaning that it can be any of $G$ and $B$.
  The left two columns show reputations, and the third column is the action $A$
  prescribed by the action rule. The fourth column indicates the reputation
  assigned to the donor who executed the action $A$, and the last column shows
  the reputation resulting from the other action $\lnot{A}$.
  The dagger ($\dagger$) means that the action is either $C$ or $D$ depending on
  the assignment rule, so it is $C$ if and only if $R(B,B,C) =
  G$ and $R(B,B,D) = B$.
  }
  \label{tab:leading_eight}
  \begin{tabular}{|cc|ccc|} \hline
    donor & recipient & prescribed action $A$ & reputation for $A$ & reputation for $\lnot{A}$ \\ \hline
    G & G & C & G & B \\
    G & B & D & G & $\ast$ \\
    B & G & C & G & B \\
    B & B & $\dagger$ & $\ast$ & $\ast$ \\ \hline
  \end{tabular}
\end{table}

To describe a norm concisely, we hereafter use a notation composed of five
characters separated by semi-colons such as $GB{:}DG{:}N$.
The first two characters denote the reputations of a donor and a recipient,
respectively. In this example, the donor's reputation is $G$, and the recipient's reputation
is $B$. The third character denotes the prescribed action, and the fourth
character means the reputation that the donor obtains by following the
prescription. Finally, the last character denotes the donor's new reputation
when choosing the opposite action. The above example is thus interpreted as
follows:
``When a $G$-donor meets a $B$-recipient, the donor should defect. He or she
gets $G$ if following the prescription, and $N$ otherwise.''
In addition, square brackets $[\dots]$ and a wildcard $\ast$ are used to
indicate a set of prescriptions.
For instance, $GB{:}D[NG]{:}B$ means a set of prescriptions, according to which a
$G$-donor should defect against a $B$-recipient. By defecting, the donor earns
either $N$ or $G$. Otherwise, the donor's reputation becomes $B$.
By this notation, the leading eight can be characterized
by the following prescriptions:
\begin{subequations}
\begin{align}
    &GG{:}CG{:}B \label{eq:GGCGB}\\
    &GB{:}DG{:}\ast \label{eq:GBDG}\\
    &BG{:}CG{:}B \label{eq:BGCG}.
\end{align}
\label{eq:leading_eight}
\end{subequations}
%where we have ignored the case of both having $B$
%because the majority of players have $G$ reputation, i.e. $h_G^{\ast} \approx 1$.
In \track{a stationary state}, the fraction of $B$-players $h_B^{\ast}$ and the defection level $p_d$ scale as $O(\mu)$.
This is because players get $B$-reputation only by implementation or assignment errors,
whereas a $B$-player can almost always recover reputation by meeting a $G$-player.
One can obtain the scaling of $h_B^{\ast}$ from the master equation of $h_B$ near \track{the stationary state} ($h_B \ll h_G$) in the following form:
\begin{equation}
    \frac{d}{dt} h_B \propto -h_G h_B + O(\mu),
\end{equation}
where the first term on the right-hand side means the rate of change from $B$ to $G$ represented by Eq.~(\ref{eq:BGCG}), and the last term represents the opposite caused by error.
We note that all the wildcards in Table~\ref{tab:leading_eight} are prescribed for the events that happen with probability smaller than $O(\mu)$.
For instance, a $B$-donor meets another $B$-recipient with probability $O(\mu^2)$.
However, such an event is so rare that the prescriptions for these events remain arbitrary within the leading eight.
In other words, we can understand the working mechanism of the leading eight by investigating
the events that occur with probability $\gtrsim O(\mu)$.

Here, it is worth pointing out that the leading eight are
the only CESS's in the binary-reputation model.
If any of the prescriptions in Eq.~(\ref{eq:leading_eight}) is missing, the norm is no longer a CESS.
For instance, without the justification of punishment [Eq.~\eqref{eq:GBDG}],
the norm is essentially equivalent to Image Scoring,
which cannot sustain stable cooperation because
those who punish a $B$-player also lose good reputation, making $h_B^{\ast}$
greater than $O(\mu)$.

\subsection*{CESS's in the ternary-reputation model}

\begin{figure}
\begin{center}
\includegraphics[width=0.6\textwidth]{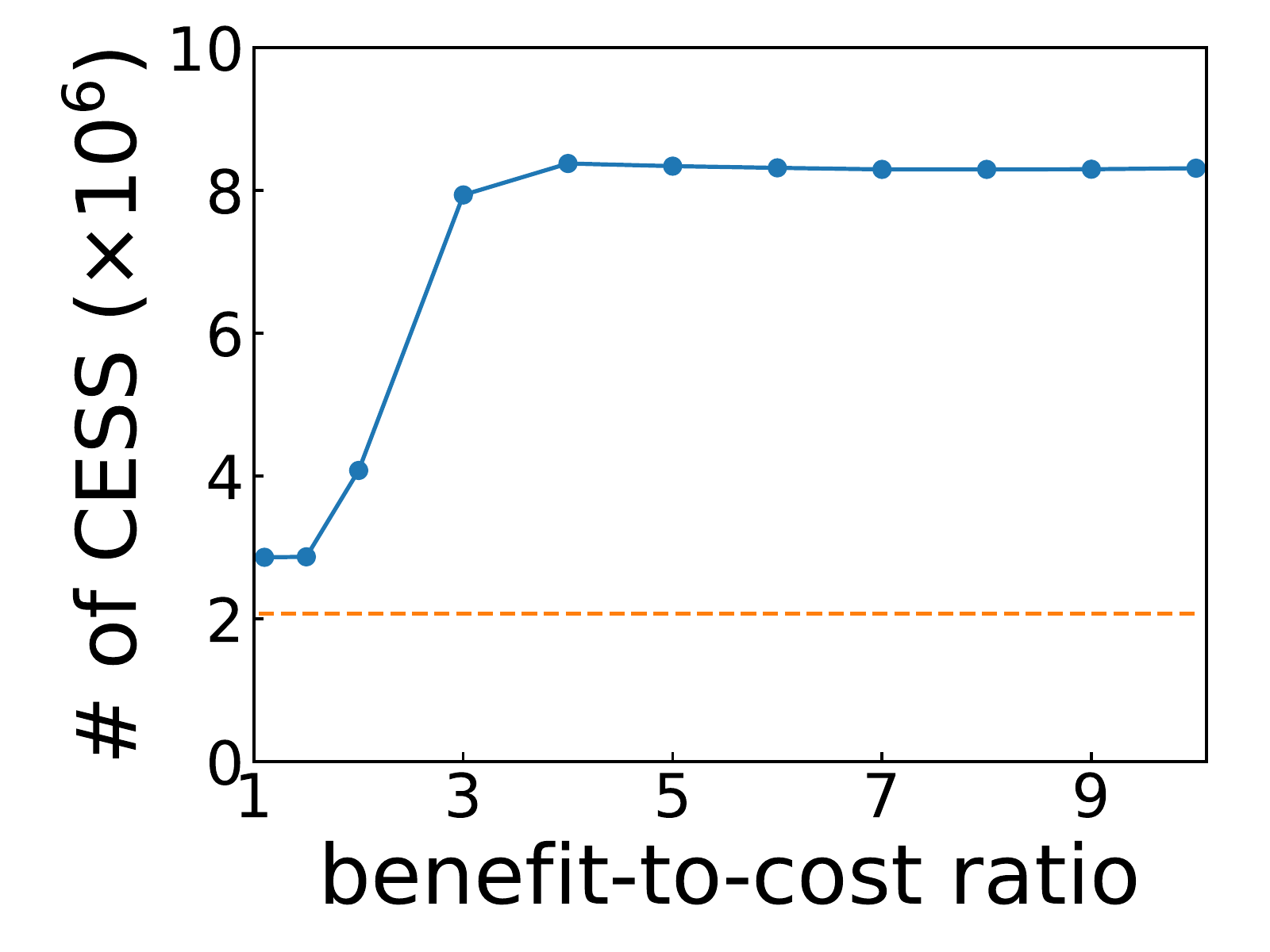}
\end{center}
\caption{
Number of the CESS's for various values of $b/c$ when $\mu_a = \mu_e = 10^{-3}$ and $p_c^{\rm th} = 0.99$.
\track{CESS's are calculated for $b/c = 1.1, 1.5, 2, 3, \dots, 10$.}
The horizontal dashed line indicates the number of the core set, which is
defined as the common subset of CESS's for \track{these} values of $b/c$.%\track{\st{ in $[1.1, 10]$}}.
}
\label{fig:num_ess}
\end{figure}

We exhaustively enumerated all the norms to find CESS's in the ternary model.
The number of CESS's is shown as a function of $b/c$ in Fig.~\ref{fig:num_ess}.
As shown in this figure, the number tends to increase in a stepwise fashion with $b/c$,
indicating the existence of norms that qualify as CESS's only for a certain range of $b/c$.
Let us define the ``core'' set as norms that are evolutionarily stable
within a reasonable range of $b/c$, say, $[1.1, 10]$.
In other words, the core set is the common subset of the discovered CESS's.
Of course, even
the norms in the core set may be evolutionarily unstable if $b/c$ is extremely
large or close to unity, but such edge cases were excluded from consideration.
Our core set contained \track{$2,067,861$} CESS's in total, and we examined this set.
\track{
Note that the size of the core set is smaller than the number of CESS's for the
lowest $b/c$. Furthermore, the number of CESS's does not increase monotonically
as $b/c$ grows in Fig.~\ref{fig:num_ess}. Such behaviour implies the existence
of nontrivial social norms that are CESS's for a certain value of $b/c$ but not
when $b/c$ takes a higher value.
For example, if a player may lose good reputation by punishing an ill-reputed
player, it could be better to overlook him or her than to inflict costly
punishment as long as $b/c$ is high enough.
}

\begin{figure}
\begin{center}
\includegraphics[width=0.9\textwidth]{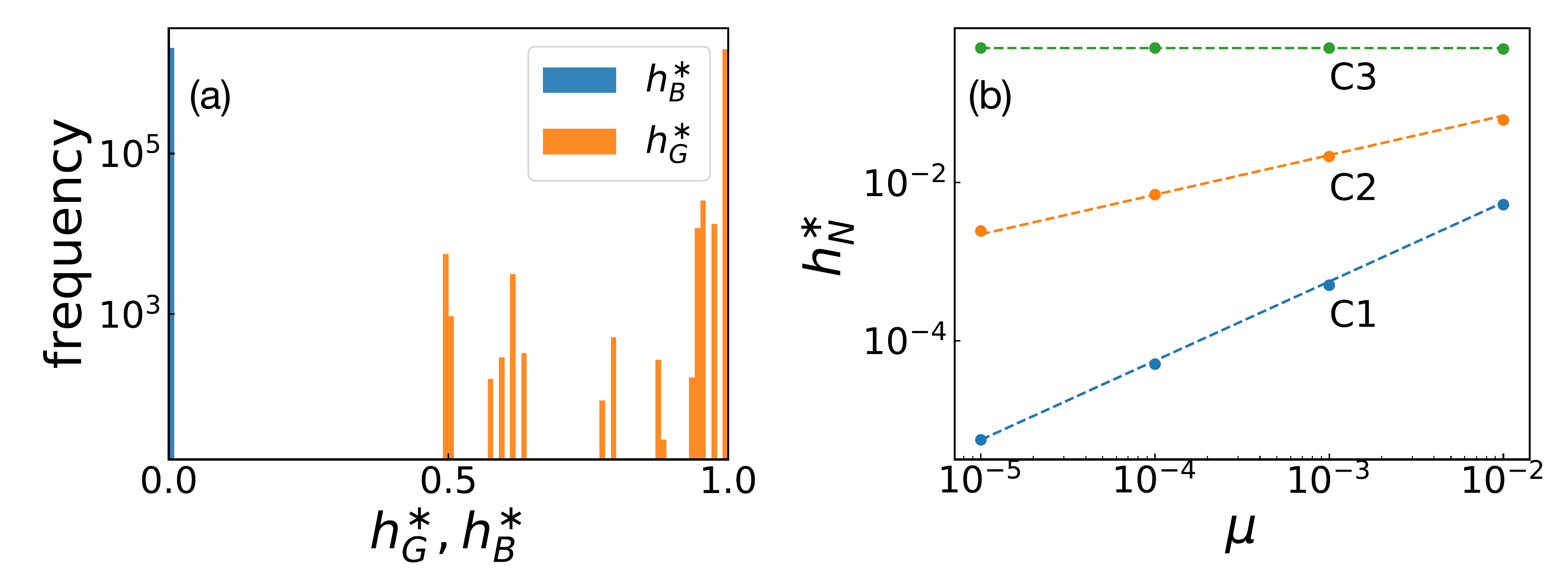}
\end{center}
\caption{
(a) Frequency of the \track{stationary-state} fractions $h_G^{\ast}$ and $h_B^{\ast}$,
respectively, for the core set. The vertical axis is on a logarithmic scale.
(b) Scaling relations between $h_N^{\ast}$ and $\mu$.
The norms used in this plot are taken from the first rows of Tables~\ref{tab:c1_strategies}-\ref{tab:c3_strategies}.
The dashed lines have power-law exponents $0$, $1/2$, and $1$, respectively.
}
\label{fig:h_dist}
\end{figure}

Figure~\ref{fig:h_dist}(a) shows the distributions of $h_B^{\ast}$ and $h_G^{\ast}$
for the norms in the core set.
The plot of $h_N^{\ast}$ has been omitted because $h_N^{\ast} = 1 - h_B^{\ast} - h_G^{\ast}$.
The figure shows $h_{B}^{\ast} \approx 0$ for all the cases,
and we numerically verified that $h_B^{\ast} \sim O(\mu)$ for $\mu \ll 1$.
However, whereas most norms have $h_{G}^{\ast} \approx 1$ similarly to the leading eight,
we found a small but non-negligible amount of norms for which
$h_{G}^{\ast}$ is significantly smaller than unity, indicating the existence of
CESS's having different working mechanisms from those of the leading eight.

To understand CESS's systematically, we first classified them according to how much $G$-players exist in \track{the stationary state}.
For the leading eight, the majority of players have reputation $G$, i.e. $h_G^{\ast} \sim O(1)$ and $h_B^{\ast} \sim O(\mu)$, and mutual cooperation is formed by these $G$-players.
For some of the ternary strategies, on the other hand, not only $G$ but $N$ may occupy a significant fraction of the population as shown in Fig.~\ref{fig:h_dist}(a).
Depending on the scaling behaviours of $h_N^{\ast}$ as $\mu \to 0$, we found that the CESS's in the core set are classified into the following three types:
\begin{itemize}
  \item Type C1: Norms with $h_N^{\ast} = O(\mu)$.
  \item Type C2: Norms with $h_N^{\ast} = O(\mu^{1/2})$.
  \item Type C3: Norms with $h_N^{\ast} = O(\mu^0)$.
\end{itemize}
Figure~\ref{fig:h_dist}(b) shows examples of the scaling relations between $h_N^{\ast}$ and $\mu$ for three norms, one in each class.
In this way, despite the considerable differences at the prescription level, the vast number of CESS's can be categorized into three well-defined classes unambiguously.

We confirmed that all the core CESS's, as in the leading eight, commonly have mechanisms
to punish defectors and to recover their reputations from erroneous actions.
However, we also observed a couple of variants in their ways of punishment and recovery.
After classifying the norms into the above three types, we further classified them
according to how players conducted punishment and recovery, which we call punishment and recovery patterns.

What do we mean by punishment patterns?
With the leading eight, the majority $G$-players punish $B$-players by defecting
against them, and those who inflicted punishment keep $G$-reputation after their
punishment. Namely, the punishment is justified.
However, this is not always the case for some of the ternary CESS's, under which
a punishing player's reputation does change.
We thus classify the norms into those with full justification (P1) and those with partial justification (P2).

% For the CESS's, those who get $B$ reputation are punished by
% the majority players (which is $G$-players for C1 and C2, and a group of $G$-players and $N$-players for C3).
% The punishment pattern indicates how players punish those who defected.
% For instance, a $G$-player who defected against $B$-player gets $N$ under some of the norms.

% There are also several types in the ways how $B$-players recover their reputations.
% For instance, in some C1 norms, $B$-players first become $N$ before recovering $G$-reputation
% while other norms allow them to directly go back to $G$.
% Norms are classified according to these patterns as shown later.

Likewise, a recovery pattern means the way that $B$-players recover their
reputations.
With the CESS's, a $B$-player can restore reputation by making an apology.
In the leading-eight community, a $B$-player can immediately return to $G$ after
cooperating with a $G$-player, and such norms that allow instantaneous recovery
are labeled as R1.
However,
R1 is not the unique recovery pattern in the case of the ternary reputation
because some norms allow $B$-players to only gradually recover their reputation,
which we call R2.

\begin{figure}
\begin{center}
\includegraphics[width=0.9\textwidth]{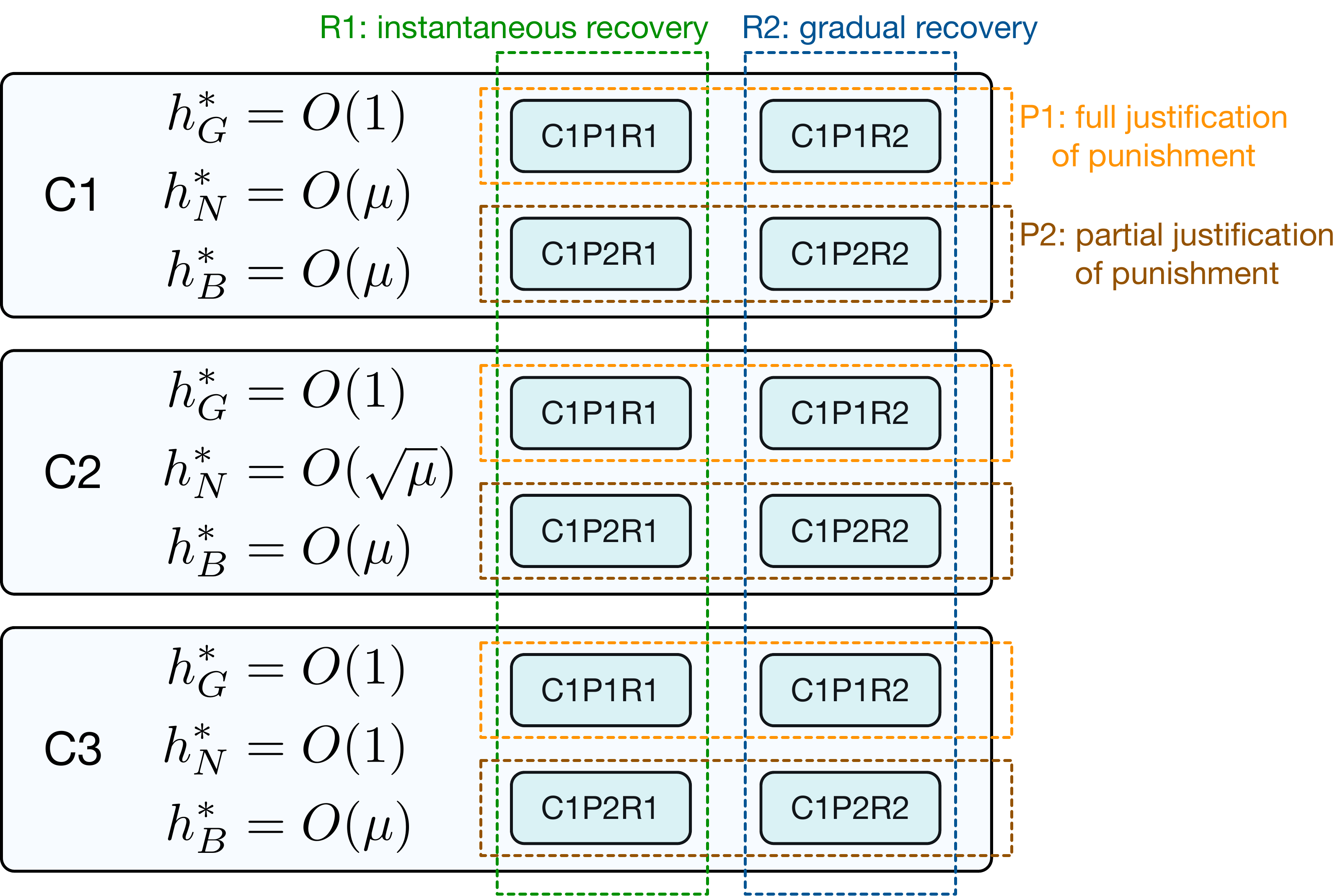}
\end{center}
\caption{
The taxonomy of the CESS's.
They are classified into C1, C2, and C3 types according to the scaling of $h_N^{\ast}$.
Each class is further categorized into four subclasses.
Norms that fully (partially) justify punishment are labeled as P1 (P2).
They are also categorized according to whether $B$-players are allowed to recover their reputation instantaneously (R1) or gradually (R2).
The leading eight correspond to C1P1R1 in which $N$ is totally irrelevant, or to
C3P1R1 when $G$ and $N$ merge into a single reputation.
}
\label{fig:taxonomy}
\end{figure}

In summary, the CESS's are grouped into three classes and 12 subclasses according to the taxonomy shown in Fig.~\ref{fig:taxonomy}.
We will see the details in the following.

\subsection*{Details of each type}

\subsubsection*{Type C1}

The first class (C1) is the most common type.
It contains more than two million norms, which comprises \track{about} $97\%$ of the core CESS's.
With this class of norms, the majority of the players have reputation $G$ and form mutual cooperation.
In other words, the norms prescribe $GG{:}CG{:}B$ in common.
The master equations near \track{the stationary state} are approximated in the following forms:
\begin{eqnarray}
\frac{d}{dt}h_N &\propto& -h_G h_N + O(\mu)\label{eq:c1N}\\
\frac{d}{dt}h_B &\propto& -h_G h_B + O(\mu)\label{eq:c1B},
\end{eqnarray}
which mean that players with $N$ or $B$ quickly change reputation by meeting $G$-players,
the majority of the population ($h_G \approx 1$).
If $\mu \ll 1$, we can see from Eqs.~\eqref{eq:c1N} and \eqref{eq:c1B} that both $h_N$ and $h_B$ will decrease exponentially as time goes by.
In \track{the stationary state}, the population will thus end up with $h_N \sim h_B \sim O(\mu)$, sharing a high degree of similarity to the leading eight.
Nevertheless, we find some distinctions from the leading eight in the punishment and recovery patterns.
Example norms in each subclass are shown in Table~\ref{tab:c1_strategies}.

Let us first look at two different patterns, depending on which reputation is assigned to a punishing player:
\begin{itemize}
  \item Type P1: Norms with $GB{:}DG{:}\ast$,
  \item Type P2: Norms with $GB{:}DN{:}\ast$.
\end{itemize}
As we have seen in the binary-reputation model, a punishing $G$-player must not
get $B$-reputation
to keep the cooperation level high. Thus, the above two are the only possibilities.
Class P1 works similarly to the leading eight. Namely, the punishing behaviour
is fully justified, and a $G$-player can maintain the reputation.
On the other hand, P2 is a novel class that has not been reported before.
Under a P2 norm, a punishing $G$-player cannot maintain his or her original reputation but gets $N$.
Their punishment is not always justified because the resulting $N$-players are
also punished by $G$-players under some of the P2 norms.
Nevertheless, the fractions of $B$- and $N$-players remain $O(\mu)$ because of the prescription $GN{:}[CD]G{:}\ast$ which is commonly found in P2.
In other words, actions against $N$-players are always justified in P2.

C1 norms are also classified according to recovery patterns as well:
\begin{itemize}
    \item Type R1: Norms with $BG{:}CG{:}B$, % $B$-players obtain $G$-reputation after interacting with $G$ ().
    \item Type R2: Norms with $BG{:}[CD]N{:}*$ and $NG{:}[CD]G{:}*$. % Two steps are required for $B$-players to go back to $G$ ().
\end{itemize}
R1 is the most basic type, similar to the leading eight.
R2 is unique to the ternary model: It takes two steps for a $B$-player to improve reputation to $G$.
During the recovery process, a player needs to cooperate with $G$-players at least once
to ensure that defection does not pay.

\begin{table}[ht]
  \centering
  \caption{
  Examples of C1 strategies.
  The \track{stationary-state} fraction of players $(h_B^{\ast},h_N^{\ast},h_G^{\ast})$ and the cooperation level $p_c$
  for $\mu_a = \mu_e = 10^{-3}$ are shown together with their prescriptions.}
  \label{tab:c1_strategies}
  \begin{tabular}{|c|ccc|c|c|} \hline
  Type & \multicolumn{3}{c|}{Prescriptions} & $(h_B^{\ast},h_N^{\ast},h_G^{\ast})$ & $p_c$ \\ \hline
           & $BB{:}DG{:}B$ & $BN{:}DN{:}B$ & $BG{:}CG{:}N$ & & \\  % ID: 146030590244
  C1-P1-R1 & $NB{:}DB{:}N$ & $NN{:}DN{:}B$ & $NG{:}CG{:}B$ & $(0.0015,0.0005,0.9980)$ & $0.9980$\\
           & $GB{:}DG{:}N$ & $GN{:}DG{:}N$ & $GG{:}CG{:}B$ & & \\ \hline

           & $BB{:}DN{:}N$ & $BN{:}DG{:}G$ & $BG{:}CN{:}B$ & & \\ % ID: 138662168852 (C1.P1.R22)
  C1-P1-R2 & $NB{:}DG{:}G$ & $NN{:}CN{:}B$ & $NG{:}DG{:}N$ & $(0.0015,0.0020,0.9965)$ & $0.9945$\\
           & $GB{:}DG{:}N$ & $GN{:}DG{:}B$ & $GG{:}CG{:}B$ & & \\ \hline

           & $BB{:}DG{:}G$ & $BN{:}DB{:}B$ & $BG{:}CG{:}B$ & & \\ % ID: 153072243972 (C1-P21-R1)
  C1-P2-R1 & $NB{:}DG{:}N$ & $NN{:}DB{:}B$ & $NG{:}DG{:}N$ & $(0.0015,0.0020,0.9965)$ & $0.9945$\\
           & $GB{:}DN{:}N$ & $GN{:}DG{:}G$ & $GG{:}CG{:}B$ & & \\ \hline

           & $BB{:}DG{:}B$ & $BN{:}DG{:}N$ & $BG{:}DN{:}N$ & & \\ % ID: 138465867040 (C1-P21-R23)
  C1-P2-R2 & $NB{:}DN{:}N$ & $NN{:}DG{:}G$ & $NG{:}CG{:}N$ & $(0.0015,0.0035,0.9950)$ & $0.9936$\\
           & $GB{:}DN{:}N$ & $GN{:}DG{:}B$ & $GG{:}CG{:}B$ & & \\ \hline
  \end{tabular}
\end{table}

\begin{figure}
\begin{center}
\includegraphics[width=0.9\textwidth]{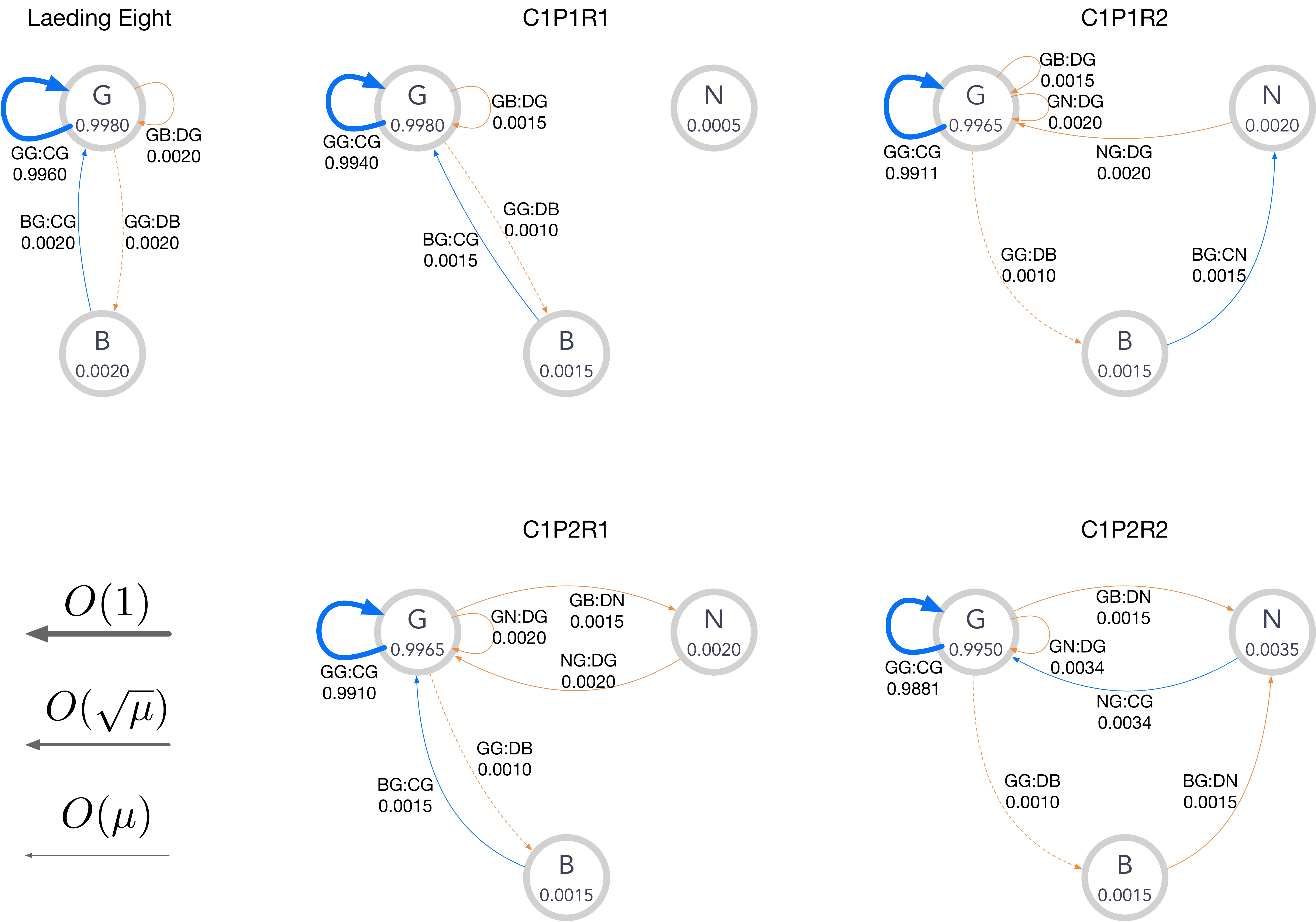}
\end{center}
\caption{
%The state transitions for the C1 norms shown in Table~\ref{tab:c1_strategies}.
Graphical representation of C1 norms.
In each node, the label of the corresponding reputation and its \track{stationary-state} fraction $h_X^{\ast}$ are shown.
An edge $XY{:}AZ$ means transition from $X$ to $Z$, where
$X$ and $Y$ are respective reputations of the donor and the recipient, $A$ is the action, and $Z$ is the new reputation assigned to the donor.
%Only edges with weights greater than a certain threshold are shown.
Only edges with weights $\gtrsim O(\mu)$ are shown for the sake of visibility.
The edges are depicted in blue (red) when the action is $C$ ($D$), and their widths indicate the amounts of the probability flow.
The dashed edges indicate the moves caused by implementation error.
The graph for the leading eight is also shown for comparison.
}
\label{fig:C1_graph}
\end{figure}

Figure~\ref{fig:C1_graph} represents state transition for the C1 norms in Table~\ref{tab:c1_strategies}.
Transitions in \track{the stationary state} [Eq.~\eqref{eq:txy_z}] are depicted as weighted edges.
The corresponding graph for the leading eight is also presented for comparison:
The thick blue self-loop at $G$ indicates that the majority of the players have $G$ with a high level of cooperation.
When an implementation error happens, the state moves to $B$ as indicated by the dashed edge $GG{:}DB$.
The other self-loop is $GB{:}DG$, which means justified punishment inflicted by $G$-players.
The remaining edge, $BG{:}CG$, corresponds to the recovery of reputation.

%The graphs for the C1 norms can be interpreted in similar ways.
The topology of the graph for the C1P1R1 norm is identical to that for the leading eight except for the unused node $N$, indicating that their working mechanisms are essentially the same.
P2 norms have a directed edge from $G$ to $N$ in common, which corresponds to $GB{:}DN$ instead of the self-loop $GB{:}DG$ in P1 norms,
and it implies reputation change caused by punishment.
Whereas punishment against $B$-players is not justified, the action against $N$-players is justified
as seen in the self-loop $GN{:}DG$ or $GN{:}CG$, which is required to keep $h_G$ high.
We can also find difference between R1 and R2 in that
R2 norms require two steps to reach $G$ from $B$, as seen from
a path $B \to N \to G$ instead of a direct edge $B \to G$.

\track{The numbers of CESS's in each class and their percentages to the entire core set are the following:
C1P1R1 has $1,057,956$ norms ($51.2\%$), C1P1R2 has $395,829$ norms ($19.1\%$), C1P2R1 has $281,586$ norms ($13.6\%$), and C1P2R2 has $269,790$ norms ($13.0\%$).
}

\subsubsection*{Type C2}

The second class contains $51,363$ norms, and this number is much smaller than C1.
In this class, we have $h_N^{\ast} \sim O(\sqrt{\mu})$, which is small but significantly greater than $O(\mu)$.
Near \track{the stationary state}, the leading terms of the master equations are written as follows:
  \begin{eqnarray}
      \frac{d}{dt}h_N &\propto& -h_N^2 + O(\mu)\label{eq:c2N}\\
      \frac{d}{dt}h_B &\propto& -h_G h_B + O(\mu).\label{eq:c2B}
  \end{eqnarray}
The dynamics of $h_B$ are the same as C1: It decreases exponentially fast, and its stationary value is of $O(\mu)$.
As for $h_N$, if $\mu$ is negligibly small, $dh_N/dt \propto -h_N^2$ is solved by $h_N \sim 1/t$.
Therefore, although $h_G$ approaches $100\%$, this process is slow with a diverging time scale.
In \track{the stationary state}, Eq.~\eqref{eq:c2N} implies $h_N \sim O(\mu^{1/2})$, which is significantly greater than $O(\mu)$.
Differently from C1, Eq.~\eqref{eq:c2N} shows that a $G$-player's encounter with an $N$-player does not decrease the total number of $N$-players because their reputations are either preserved or swapped by this event.
The fraction $h_N$ decreases mainly when two $N$-players meet, making the difference from C1.

Cooperation is prescribed not only between two $G$-players but between a $G$-player and a $N$-player,
i.e., $GG{:}CG{:}B$, $GN{:}C[GN]{:}B$, $NG{:}C[GN]{:}B$.
When two $N$-players meet, however, they may defect under some of the C2 norms
because such events occur with probability of $O(\mu)$ and do not decrease the cooperation level significantly.

As in C1, C2 norms can be further classified according to punishment patterns as
follows:
\begin{itemize}
    \item Type P1: Norms with $GB{:}DG{:}\ast$
    \item Type P2: Norms with $GB{:}DN{:}\ast$,
\end{itemize}
where P1 means full justification, whereas P2 does partial one.
We can also classify C2 norms according to recovery patterns:
\begin{itemize}
    \item Type R1: Norms with $BG{:}CG{:}B$
    \item Type R2: Norms with $BG{:}CN{:}B$,
\end{itemize}
where
R1 and R2 correspond to the instantaneous and gradual recovery processes, respectively.

\begin{figure}
\begin{center}
\includegraphics[width=0.9\textwidth]{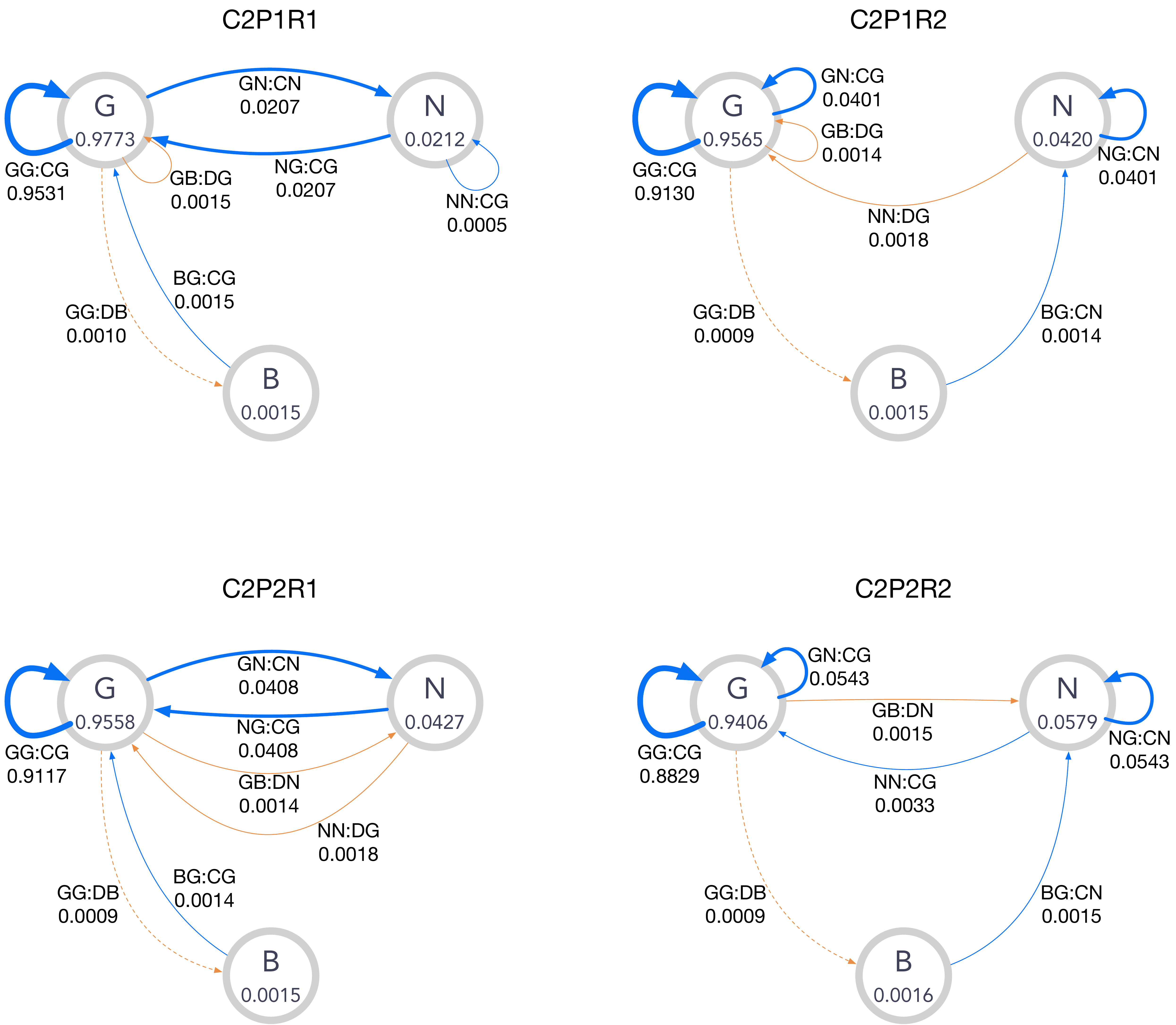}
\end{center}
\caption{
The state transitions for the C2 norms shown in Table~\ref{tab:c2_strategies}.
The notations of the graphs are the same as those in Fig.~\ref{fig:C1_graph}.
}
\label{fig:C2_graph}
\end{figure}

Examples of C2 norms are shown in Table~\ref{tab:c2_strategies},
and their state-transition graphs are depicted in Fig.~\ref{fig:C2_graph}.
The difference between these subclasses is clear:
The graphs for P1 have self-edges $GB{:}DG$ in common, whereas those for P2 have edges from $G$ to $N$ ($GB{:}DN$).
Similarly, the graphs for R1 have edges from $B$ to $G$ ($BG{:}CG$) whereas those for R2 have a path $B \to N \to G$.

% Thus, cooperation must be prescribed not only between two $G$-players but between a $G$-player and a $N$-player,
% i.e., $GG{:}CG{:}B$, $GN{:}C[GN]{:}B$, $NG{:}C[GN]{:}B$ are commonly prescribed.
% At first glance, it looks like the leading eight if we merge $G$ and $N$ into a single reputation.
% However, there is an interesting difference:
% When a $N$-donor plays against another $N$-recipient, the donor may defect in this class.
% The probability of an encounter between two $N$-players is $O(\mu)$ as $h_N^{\ast} = O(\sqrt{\mu})$
% therefore the cooperation level is kept high enough even if defection is prescribed for this event.
% While the total cooperation level is high enough, a $N$-player has a significantly high chance to be
% defected therefore having $G$-reputation is significantly beneficial than $N$-reputation from a single player's perspective.

% The mechanism behind the scaling $h_N^{\ast} = O(\sqrt{\mu})$ is explained as follows.
% In C2, the prescription between $G$ and $N$ are limited to two cases:
% ($GG{:}CG{:}B, GN{:}CG{:}B, NG{:}CN{:}B, NN{:}[CD]G{:}\ast$) or
% ($GG{:}CG{:}B, GN{:}CN{:}B, NG{:}CG{:}B, NN{:}[CD][GB]{:}\ast$).
% XXX [TODO] XXX

\begin{table}[ht]
  \centering
  \caption{
  Examples of C2 norms. The \track{stationary-state} fraction of players $(h_B^{\ast},h_N^{\ast},h_G^{\ast})$ and the cooperation level $p_c$ for $\mu_a = \mu_e = 10^{-3}$ are shown together with their prescriptions.
  }
  \label{tab:c2_strategies}
  \begin{tabular}{|c|ccc|c|c|} \hline
  Type & \multicolumn{3}{c|}{Prescriptions} & $(h_B^{\ast},h_N^{\ast},h_G^{\ast})$ & $p_c$ \\ \hline
           & $BB{:}DN{:}N$ & $BN{:}CN{:}B$ & $BG{:}CG{:}B$ & & \\ % ID: 141151461302 C22-P1-R1
  C2-P1-R1 & $NB{:}DG{:}G$ & $NN{:}CG{:}B$ & $NG{:}CG{:}B$ & $(0.0015,0.0212,0.9773)$ & $0.9985$\\
           & $GB{:}DG{:}N$ & $GN{:}CN{:}B$ & $GG{:}CG{:}B$ & & \\ \hline

           & $BB{:}DN{:}B$ & $BN{:}DG{:}N$ & $BG{:}CN{:}B$ & & \\ % ID:148413257644 C21-P1-R2
  C2-P1-R2 & $NB{:}CG{:}B$ & $NN{:}DG{:}G$ & $NG{:}CN{:}B$ & $(0.0015,0.0420,0.9565)$ & $0.9967$\\
           & $GB{:}DG{:}N$ & $GN{:}CG{:}B$ & $GG{:}CG{:}B$ & & \\ \hline

           & $BB{:}DG{:}N$ & $BN{:}DN{:}N$ & $BG{:}CG{:}B$ & & \\ % ID:140057476516 C22-P2-R1
  C2-P2-R1 & $NB{:}DG{:}B$ & $NN{:}DG{:}N$ & $NG{:}CG{:}B$ & $(0.0015,0.0427,0.9558)$ & $0.9966$\\
           & $GB{:}DN{:}B$ & $GN{:}CN{:}B$ & $GG{:}CG{:}B$ & & \\ \hline

           & $BB{:}CG{:}B$ & $BN{:}DN{:}B$ & $BG{:}CN{:}B$ & & \\ % ID:148948479413 C21-P2-R2
  C2-P2-R2 & $NB{:}DB{:}B$ & $NN{:}CG{:}B$ & $NG{:}CN{:}B$ & $(0.0016,0.0579,0.9406)$ & $0.9983$\\
           & $GB{:}DN{:}G$ & $GN{:}CG{:}B$ & $GG{:}CG{:}B$ & & \\ \hline
  \end{tabular}
\end{table}

\track{The numbers of CESS's in each class and their percentages to the entire core set are the following:
C2P1R1 has $13,275$ norms ($0.64\%$), C2P1R2 has $15,543$ norms ($0.75\%$), C2P2R1 has $10,659$ norms ($0.52\%$), and C2P2R2 has $11,886$ norms ($0.57\%$).
}

\subsubsection*{Type C3}

Finally, we categorize \track{$11,337$} norms into the third class (C3).
Differently from C1 and C2, the stationary fraction $h_N^{\ast}$ remains finite even when $\mu \to 0$.
As a result, most players end up with either $G$ or $N$.
Mutual cooperation is formed between $G$- and $N$-players, whereas those who defect receive reputation $B$.
The fraction of $B$-players is a small quantity of $O(\mu)$ because one can easily escape from $B$-reputation by meeting $G$- or $N$-players.
We also obtained asymptotic dynamics of $h_G$ and $h_N$ as shown in Methods.

Similarly to the other classes, a finer classification of C3 norms according to
their punishment patterns can be defined as follows:
\begin{itemize}
    \item Type P1: Norms with $GB{:}D[GN]{:}{\ast}$ and $NB{:}D[GN]{:}{\ast}$,
    \item Type P2: Norms with ($GB{:}DB{:}{\ast}, NB{:}D[GN]{:}{\ast}$) or ($GB{:}D[GN]{:}{\ast}, NB{:}DB{:}{\ast}$).
\end{itemize}
As before, punishment is fully justified in P1.
P2 means partial justification in the sense that only punishment inflicted by $G$-players is justified, but not the one by $N$-players, or the other way around.
C3 norms are classified according to recovery patterns as well:
\begin{itemize}
    \item Type R1: Norms with $BG{:}C[GN]{:}B$ and $BN{:}C[GN]{:}B$,
    \item Type R2: Norms with $(BN{:}DB{:}B, BG{:}C[GN]{:}B)$ or $(BN{:}C[GN]{:}B, BG{:}DB{:}B)$.
\end{itemize}
R1 is similar to the leading eight, and R2 is unique to the ternary-reputation model because
$B$-players cooperate only with either $G$- or $N$-players and recover
reputation.
Thus, on average, more than one step is required to return to the original
reputation state if it is lost by mistake.

\begin{figure}
\begin{center}
\includegraphics[width=0.9\textwidth]{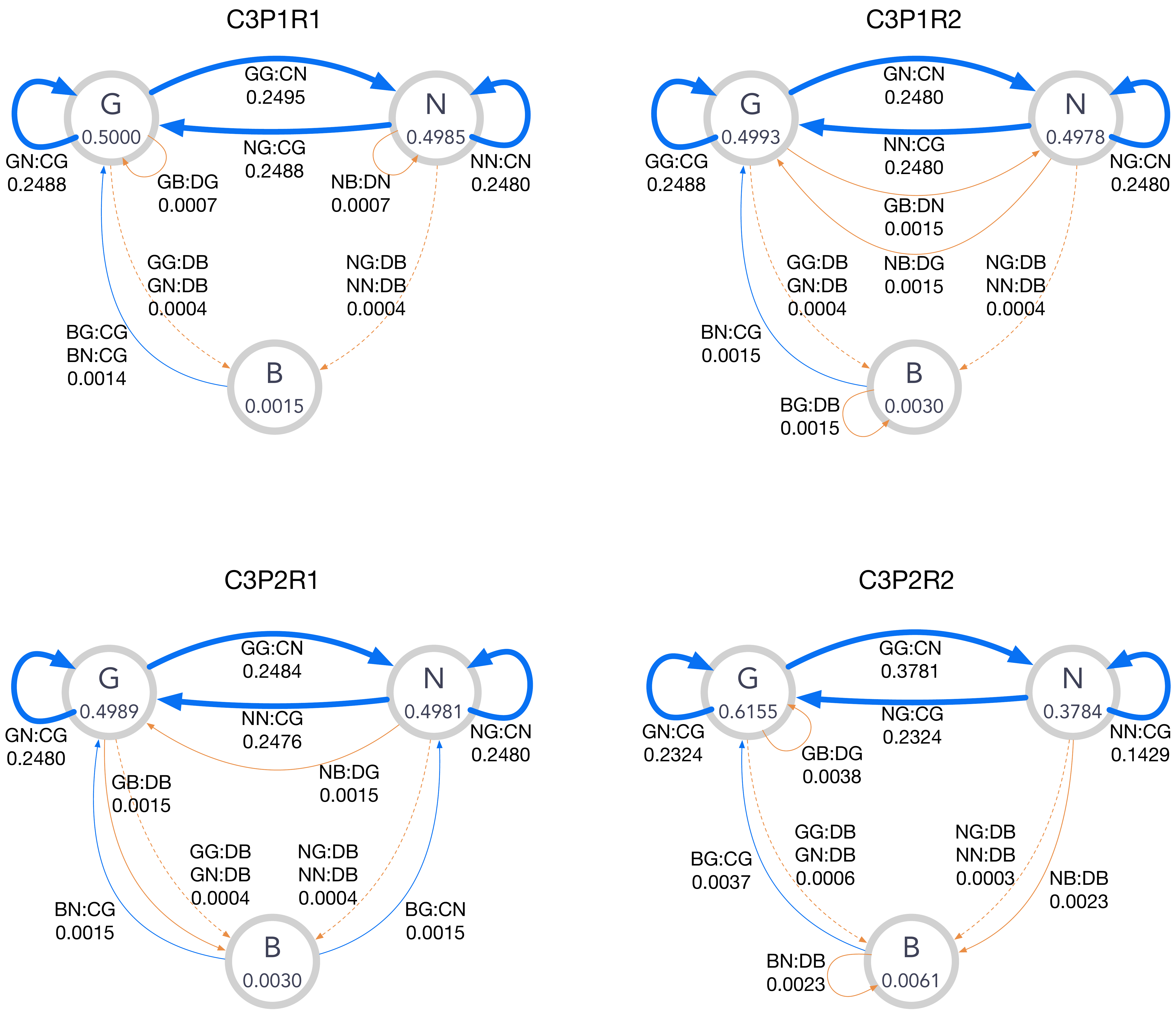}
\end{center}
\caption{
The state transitions for the C3 norms shown in Table~\ref{tab:c3_strategies}.
The notations of the graphs are the same as those in Fig.~\ref{fig:C1_graph}.
}
\label{fig:C3_graph}
\end{figure}

Examples of C3 norms as well as their state-transition graphs are shown in Table~\ref{tab:c2_strategies} and in Fig.~\ref{fig:C3_graph}.
The graph for C3P1R1 is equivalent to that for the leading eight if $G$ and $N$ merge into one.
Whereas P1 norms have no solid edges to $B$, P2 norms have a solid edge to $B$ either from $G$ or $N$, indicating that punishment is not always justified in P2 norms.
No self-loop exists around $B$ in the R1 graphs whereas those for the R2 graphs have a self-loop $BN{:}DB$ or $BG{:}DB$, indicating that $B$-players cannot always escape from $B$.

\track{The numbers of CESS's in each class and their percentages to the entire core set are the following:
C3P1R1 has $5,199$ norms ($0.25\%$), C3P1R2 has $4,413$ norms ($0.21\%$), C3P2R1 has $1,593$ norms ($0.08\%$), and C3P2R2 has $132$ norms ($0.01\%$).
}

\begin{table}[ht]
  \centering
  \caption{
  Examples of C3 strategies. The \track{stationary-state} fraction of players $(h_B^{\ast},h_N^{\ast},h_G^{\ast})$ and the cooperation level $p_c$ for $\mu_a = \mu_e = 10^{-3}$ are shown together with their prescriptions.
  }
  \label{tab:c3_strategies}
  \begin{tabular}{|c|ccc|c|c|} \hline
  Type & \multicolumn{3}{c|}{Prescriptions} & $(h_B^{\ast},h_N^{\ast},h_G^{\ast})$ & $p_c$ \\ \hline
           & $BB{:}DN{:}N$ & $BN{:}CG{:}B$ & $BG{:}CG{:}B$ & & \\ % ID: 83184194486
  C3-P1-R1 & $NB{:}DN{:}G$ & $NN{:}CN{:}B$ & $NG{:}CG{:}B$ & $(0.0015,0.4985,0.5000)$ & $0.9985$\\
           & $GB{:}DG{:}G$ & $GN{:}CG{:}B$ & $GG{:}CN{:}B$ & & \\ \hline

           & $BB{:}DG{:}N$ & $BN{:}CG{:}B$ & $BG{:}DB{:}B$ & & \\ % ID: 139969896882 C3-P1-R22
  C3-P1-R2 & $NB{:}DG{:}B$ & $NN{:}CG{:}B$ & $NG{:}CN{:}B$ & $(0.0030,0.4978,0.4993)$ & $0.9955$\\
           & $GB{:}DN{:}B$ & $GN{:}CN{:}B$ & $GG{:}CG{:}B$ & & \\ \hline

           & $BB{:}DG{:}B$ & $BN{:}CG{:}B$ & $BG{:}CN{:}B$ & & \\ % ID:80924798902 C3-P22-R1
  C3-P2-R1 & $NB{:}DG{:}B$ & $NN{:}CG{:}B$ & $NG{:}CN{:}B$ & $(0.0030,0.4981,0.4989)$ & $0.9970$\\
           & $GB{:}DB{:}B$ & $GN{:}CG{:}B$ & $GG{:}CN{:}B$ & & \\ \hline

           & $BB{:}DG{:}B$ & $BN{:}DB{:}B$ & $BG{:}CG{:}B$ & & \\ % ID:83191630772 C3-P21-R21
  C3-P2-R2 & $NB{:}DB{:}B$ & $NN{:}CG{:}B$ & $NG{:}CG{:}B$ & $(0.0061,0.3784,0.6155)$ & $0.9916$\\
           & $GB{:}DG{:}G$ & $GN{:}CG{:}B$ & $GG{:}CN{:}B$ & & \\ \hline
  \end{tabular}
\end{table}

\subsection*{Counter-intuitive examples}

Here we show a couple of norms with interesting differences from the leading eight
and discuss why they nevertheless qualify as CESS's with the ternary reputation.

The first example is ``unfair'' punishment, which is observed from the C1P2R1 norm
in Table~\ref{tab:c1_strategies} and Fig.~\ref{fig:C1_graph}.
Under this norm, a punishing player does not maintain $G$ but gets $N$,
and the player is punished by a $G$-player in the subsequent round.
In other words, a player who initially had $G$-reputation gets unfairly punished
after the unfortunate encounter with a $B$-player although the player has accurately followed the prescription.
This unfairness is never observed in the leading eight
because $B$ cannot be assigned to a punishing player to keep $h_B^{\ast} \sim O(\mu)$.
However, when the reputation is not binary, a norm without the full justification can be a CESS
because both $h_B^{\ast}$ and $h_N^{\ast}$ can be of $O(\mu)$, justifying
the action against $N$-players ($GN{:}[CD]G{:}{\ast})$.

In the second example, we see peculiar behaviour of ``making an apology by
defecting.''
Such behaviour is observed from the C1P1R2 example in Table~\ref{tab:c1_strategies} and Fig.~\ref{fig:C1_graph}.
A player needs two steps ($B \to N \to G$) to return to $G$ once he or she gets $B$.
First, a $B$-player must cooperate with a $G$-player to become $N$.
Then, the $N$-player must defect, not cooperate, against a $G$-player to become $G$,
which goes against our common sense.
This counter-intuitive apology is not possible with the binary reputation
because such a norm would allow a constantly defecting player to be better off than the rest of the population.
However, when more than two kinds of reputation are available, this is not the case.
A norm can be a CESS as long as cooperation is prescribed at least once in the course of apology, but not necessarily twice or more,
because a single move of cooperation is enough to compensate for the defection.

Another interesting behaviour is found in C2P2R1 norm in Table~\ref{tab:c2_strategies} and in Fig.~\ref{fig:C2_graph},
where ``inequality'' among cooperators spontaneously emerges.
As shown above, C2 norms are characterized by the fact that most players have
$G$ except a small fraction $O(\sqrt{\mu})$ of $N$ players in \track{the stationary state}.
Those players form mutual cooperation, but $N$-players defect against each other under some C2 norms.
Such an event occurs with probability $O(\mu)$, thus negligible at the societal level, but it makes a significant difference from a player's perspective.
That is, although $G$-players almost surely receive cooperation from the community,
$N$-players do not benefit from $N$-players, yielding the drop of the individual cooperation level by $O(\sqrt{\mu})$.
This inequality becomes significant as $\mu$ grows.
For instance, when $\mu_e = \mu_a = 0.05$, we see that $h_N^{\ast} \approx 0.22$, and more than $20\%$ of the population
suffers such a loss.
Again, the drop of the cooperation level
is still acceptable at the society level, and this C2 norm qualifies as a CESS.

\section*{Summary and Discussion}

Although reputation in our real life is not always distinguished between good and bad,
most previous works have accepted an idealized assumption of binary reputation.
Some researchers have attempted to go beyond the binary reputation~\cite{nowak1998evolution,nakamura2011indirect,berger2011learning,tanabe2013indirect, berger2016stability,clark2020indirect,schmid2021unified}, and the motivation behind the ternary reputation in Tanabe et al.~\cite{tanabe2013indirect} is close to ours.
However, they studied only second-order norms and assumed an ordinal relationship among reputations to limit the number of norms to $512$.
By considering the third-order assessment rules, this study naturally takes into account
the Self strategy~\cite{leimar2001evolution} as well as all the second-order
norms (see Methods).
One may extend the binary system by representing reputation as integer
values~\cite{nowak1998evolution,clark2020indirect,berger2016stability}.
% A recent study by Lee et al.~\cite{lee2021local} assumes continuous values as reputation.
% These studies inherently assume ordinal relationships between reputations,
% which simplifies the analysis and interpretation but significantly limit the generality.
% On the other hand, we fully studied the third-order norms without
% imposing any ordinal relationships a priori for the first time.
% Our results include second-order norms as a subset, which are shown in Appendix.
% By considering the third-order assessment, this study takes into account Self strategy~\cite{leimar2001evolution},
% which is known to be a strong strategy, as a competitor.
The other extreme is to regard reputation as a continuous variable.
This approach makes it possible to use analytic tools,
but it has its own limitations because
we can only examine a small neighborhood of the existing cooperative
norm~\cite{lee2021local}.
% These previous studies have inherently assumed ordinal relationships between reputations,
% which simplifies the analysis but may limit the generality.
% On the other hand, we fully studied the third-order norms without
% imposing any ordinal relationships a priori for the first time.
Little is known about the consequences of these simplifying assumptions:
For example, some lessons from indirect reciprocity might be due to the oversimplification.
We should ask what are the fundamental properties that are preserved irrespective
of the complexity of the reputation system to sustain cooperation.
To address this question, we comprehensively searched for the CESS pairs of
assignment and action rules with ternary reputations to compare with the leading eight.
From more than thirty billion possibilities, we filtered out ``core'' CESS norms that constitute the counterpart of the well-known leading eight.

The result shows that the previous conclusions drawn from the binary- and continuous-reputation models do not fully capture the various possibilities of CESS's.
For example, under a certain norm, a player may lose $G$-reputation even when he
or she has obeyed all the prescriptions of the norm, something unimaginable in the leading eight.
Another norm requires an ill-reputed player to defect against the community to gain a better reputation.
%Or, from a dynamical perspective, it is the case for certain norms that
% cooperative equilibrium can be approached only through erroneous $B$-reputation (the first line of C3 in Table~\ref{tab:limit}).
This observation suggests that a population may well achieve cooperation by actively making use of reputations far from $G$, a possibility that has been ignored by the linear stability analysis for the continuous model. Put differently, the viewpoint of this work is that $N$ does not necessarily mean `less bad' but functions in its own way. Based on this idea, we explored the strategy space in full without imposing any ordinal relationship {\it a priori}.
\track{Indeed, we have observed cases where $N$ cannot be interpreted
as the middle reputation between $G$ and $B$.
For instance, in some of C1P2R1 norms, a $G$-player who punished a $B$-player
gets $N$ and the $N$-player can defect against $G$-players while receiving
cooperation from $G$-players, which implies that $N$ is deemed better than the
majority's reputation $G$.
Such a CESS would not be found if we assumed an ordinal relationships among the
three reputations.
}

It is still instructive to compare our findings with the leading eight. As we have seen above, the leading eight have the following characteristics:
(i) Maintenance of cooperation, (ii) Identification of defectors,
(iii) Punishment and justification of punishment, (iv) Apology and forgiveness.
Overall, these characteristics are shared with the ternary CESS's,
indicating key features universally required for general reputation systems.
However, some of the above characteristics are relaxed when we go beyond the binary assumption.
First, cooperation is not always maintained among a single type of players
but among multiple types, as seen in C2 and C3.
Second, partial justification of punishment is allowed:
A player who inflicted punishment does not always keep the original reputation,
differently from the leading eight.
Third, forgiveness may be non-instantaneous:
Instead of being forgiven right after cooperating with $G$ as in the leading eight,
it may take some steps for a $B$-player to recover his or her reputation.

%Because of the relaxed requirements to be a CESS,
%we found several interesting norms that are absent in the binary-reputation model.
%For instance, as we see in the example of C1-P2-R1,
%there is a case that those who punished a $B$-player
%are assigned a worse reputation $N$ and will be punished in a later round
%even though they sincerely followed the prescription of the norm.
%Another interesting example is found in C1-P1-R2.
%$B$-players need to defect against $G$-players
%in order to recover their reputations thus
%defection is the way of showing an apology under this norm.
%Another interesting case is the example of C2-P1-R2.
%In this case, $N$-players defect against each other
%even though they account for a significant amount of cooperators.
%These counter-intuitive examples appear only in the ternary reputation model,
%indicating that the conclusions drawn from the binary reputation model
%should not always be generalized to more general cases.

In summary, based on the results for the ternary-reputation system, we conjecture that
the CESS norms for general reputation models will share the common characteristics in a more relaxed form:
\begin{enumerate}
    \item Maintenance of cooperation by the majority (but not necessarily all)
    of the population.
    \item Identification of defectors.
    \item Punishment, followed by partial or full justification.
    \item Apology accompanied by gradual or instantaneous forgiveness.
\end{enumerate}
We believe these rules serve as useful guiding principles when designing
a norm with even more intricate reputations.

% In addition to the core set we have mainly investigated in this paper, there are even more diverse ESS pairs for a larger $b/c$.
% These additional pairs are not classified into $14$ classes we have shown.
% Although full characterization of these pairs is left for future research, the full list of ESS pairs and the source codes are available online~\cite{}.
% Here, we only show a few examples in Table~\ref{tab:other_strategies} to demonstrate the richness of the solutions.

Finally, we would like to stress that
the ternary-reputation model is an interesting system in its own right because the third reputation may provide additional historical information for players.
An important future direction is the study of the private reputation in a noisy environment~\cite{hilbe2018indirect,ohtsuki2007global,uchida2010effect,uchida2013effect,okada2017tolerant,okada2018solution,okada2020review,okada2020two}. Investigation into a polymorphic population in this context remains an open problem at large, and the same remark applies to
stochastic reputation dynamics~\cite{tanabe2013indirect,schmid2021unified}.
Because some of the ternary CESS's have the diverging time scales in the dynamics of $h_N$ (C2 and some C3 norms, see Table~\ref{tab:limit}),
these norms could show behaviours that are significantly different from the binary ones.

\section*{Methods}

\subsection*{Calculation of the \track{stationary-state} population}

Let $h_Z$ be the fraction of players having reputation $Z\in \{B,N,G\})$.
By construction, we always have $h_\text{sum} \equiv h_B + h_N + h_G = 1$. When
an $RP$ pair is
given, one can calculate the time evolution of $h_Z$ for a short time interval
$[t, t+\Delta t]$ in the error-free limit as follows:
Within this interval, we randomly choose a small fraction of players as donors,
and we denote their fraction as $\alpha \Delta t \ll 1$. The fraction of those
players having reputation $X$ is $ h_X(t) \alpha \Delta t$. A recipient is
assigned to each donor through random sampling, so the donor meets a recipient
with reputation $Y$ with probability $h_Y(t)$.
If everyone abides by the given $RP$ pair, we may rewrite the assignment rule as
$R(X,Y) \equiv R\left(X,Y,P\left(X,Y\right)\right)$. The inflow of $h_Z$ is thus
equal to $h_X(t)h_Y(t) \delta_{R(X,Y), Z}$, where $\delta_{i,j}$ is the
Kronecker delta, because the donor with $X$ has to interact with the recipient
with $Y$ and obtain new reputation $Z$ according to the rule $R(X,Y)$.
On the other hand, the outflow is $h_Z(t) \alpha \Delta t$
because the donors will have updated reputations other than $Z$ in general.
Thus, the time evolution of $h_Z(t)$ is given by
\begin{equation}
\begin{aligned}
h_Z(t+\Delta t) - h_Z(t) &= \alpha \Delta t \sum_{X,Y \in \{B,N,G\}}
h_X(t)h_Y(t) \delta_{R(X,Y), Z}
- \alpha \Delta t h_Z(t).
\end{aligned}
\end{equation}
Taking the limit of $\Delta t \to 0$, we have the following differential
equation:
\begin{equation}
\begin{aligned}
\frac{d}{dt}h_Z(t) &= \sum_{X,Y \in \{B,N,G\}} T_{XY\to Z} - h_Z(t),
\end{aligned}
\end{equation}
where $T_{XY\to Z} \equiv h_X(t) h_Y(t) \delta_{R(X,Y),Z}$ and $\alpha \equiv 1$
after rescaling the unit of time.

In the presence of implementation error, a donor fails to cooperate with
probability $\mu_e$. In other words, the prescribed action is correctly executed
with probability $1-\mu_e$, and the player must defects otherwise. By taking
into account the implementation error, $T_{XY \to Z}$ is thus redefined as
\begin{equation}
    T_{XY \to Z} \equiv h_X(t) h_Y(t) \left[ \left(1-\mu_e\right)\delta_{R(X,Y), Z} + \mu_e\delta_{R(X,Y,D), Z} \right].
\end{equation}
In the presence of assignment error, the donor does action $A$ and receives
correct reputation $Z$ with probability $(1-\mu_a)$, but $Z$ may be assigned by
mistake with probability $\mu_a/2$ although the assignment rule does not
prescribe $Z$.
Therefore, the probability that the donor obtains $Z$ is given as
\begin{equation}
(1-\mu_a) \delta_{R(X,Y,A), Z} + \frac{1}{2}\mu_a [1 - \delta_{R(X,Y,A), Z}] = (1-\frac{3}{2}\mu_a) \delta_{R(X,Y,A), Z} + \frac{1}{2}\mu_a.
\end{equation}
Thus, when both implementation and assignment errors may occur, $T_{XY \to Z}$
is redefined as
\begin{equation}
T_{XY \to Z} \equiv h_X(t) h_Y(t) \left\{ \left(1-\frac{3}{2}\mu_a \right) \left[ (1-\mu_e)\delta_{R(X,Y), Z} + \mu_e\delta_{R(X,Y,D), Z} \right] + \frac{\mu_a}{2} \right\}.
    \label{eq:txy_z}
\end{equation}
Note that the dynamics preserves $h_\text{sum} \equiv h_G + h_N + h_B =1$
because
\begin{equation}
\begin{aligned}
\frac{d}{dt}h_\text{sum} &= \sum_{X,Y \in \{B,N,G\}} \left( T_{XY \to G} + T_{XY
\to N} + T_{XY \to B} \right) - h_\text{sum} \\
    &= h_\text{sum}^2 - h_\text{sum} = 0.
\end{aligned}
\end{equation}

According to our numerical check, $h_Z(t)$ converges to a unique \track{stationary state},
$h_Z^{\ast} = \lim_{t \to \infty} h_Z(t)$, irrespective of the initial condition for most cases.
\track{
However, in some cases where multiple stationary states coexist,
we adopted the one obtained from the initial condition $(1/3, 1/3, 1/3)$,
regarding it as the most representative one.}
For each \track{social norm}, we obtain $h_Z^{\ast}$ by using the
fourth-order Runge-Kutta algorithm, normalizing $h_Z(t)$ by $h_\text{sum}$
each time step.

\subsection*{Calculation of the cooperation level and the payoffs}

Cooperation level $p_c$ for a resident species is defined as
\begin{equation}
p_c = \sum_{X,Y \in \{B,N,G\}} h_X h_Y \delta_{P(X,Y), C}.
\end{equation}
The payoff of a resident player $\pi_{\rm res}$ is calculated as
\begin{equation}
\pi_{\rm res} = p_c (1-\mu_e) (b-c).
\label{eq:payoff_res}
\end{equation}
Then, we calculate the dynamics of the fraction of mutant players for each reputation, $\{H_B,H_N,H_G\}$, when a small number of mutant players exist in the community.
The fraction of mutants having $Z$ reputation is updated as
\begin{equation}
\dot{H_Z} = \sum_{X,Y \in \{B,N,G\}} T_{XY \to Z}^{\rm mut} - H_Z(t),
\end{equation}
where
\begin{equation}
T_{XY \to Z}^{\rm mut} \equiv H_X(t) h_Y^{\ast} \left\{ \left(1-\frac{3}{2}\mu_a \right) \left[ (1-\mu_e)\delta_{\hat{R}(X,Y), Z} + \mu_e\delta_{ R(X,Y,D), Z} \right] + \frac{\mu_a}{2} \right\},
\end{equation}
where $\hat{R}_{XY} \equiv R\left(X,Y, \hat{P}\left(X,Y\right)\right)$ and $\hat{P}(X,Y)$ is the action rule of the mutant.
We numerically confirmed that $H_Z$ converges to a stationary value $H_Z^{\ast}$ after an initial transient period.

Using these stationary values, the probability that a mutant cooperate with a resident is
\begin{equation}
p_c^{\rm mut \to res} = \sum_{X,Y \in \{B,N,G\}} H_X h_Y \delta_{\hat{P}(XY), C},
\end{equation}
whereas its counterpart is
\begin{equation}
p_c^{\rm res \to mut} = \sum_{X,Y \in \{B,N,G\}} h_X H_Y \delta_{P(X,Y), C}.
\end{equation}
Using these, the payoff of the mutant is given as
\begin{equation}
\pi_{\rm mut} = p_c^{\rm res \to mut} (1-\mu_e) b - p_c^{\rm mut \to res} (1-\mu_e) c.
\label{eq:payoff_mut}
\end{equation}
%\track{
%\cancel{
%Comparing Eq.~(\ref{eq:payoff_res}) and Eq.~(\ref{eq:payoff_mut}), the resident
%is stable against the invasion of the mutant when}
%\begin{equation}
%\begin{cases}
%  \frac{b}{c} > \frac{p_c - p_c^{\rm mut \to res}}{p_c - p_c^{\rm res \to mut}} & (p_c - p_c^{\rm res \to mut} > 0) \\
%  \frac{b}{c} < \frac{p_c - p_c^{\rm mut \to res}}{p_c - p_c^{\rm res \to mut}} & (p_c - p_c^{\rm res \to mut} < 0) \\
%  p_c - p_c^{\rm mut \to res} < 0 & (p_c - p_c^{\rm res \to mut} = 0).
%\end{cases}
%\end{equation}
%\cancel{
%The intersection of the above ranges for all possible behavioral mutants yields the range of $b/c$ at which the resident norm is an ESS.
%}}

\subsection*{Enumeration of norms}

We enumerated all possible combinations of assignment and action rules to find every CESS's.
A supercomputer was used to deal with a large number of possibilities that amounts to $64,573,605 \times 2^{9} = 33,061,685,760$.
To speed up the calculation, we removed some of the norms that cannot be CESS's as follows:
When an assignment rule $R$ contains a case where the assigned reputation is the same for both actions, $D$ must be prescribed to be an ESS.
For instance, when $R(G,G,C) = R(G,G,D)$, a $G$ player would have no incentive to cooperate with another $G$ player.
In such a case, an action rule prescribing $C$ at $(G, G)$ cannot be an ESS because the defector gains a strictly higher payoff than the resident.
We exclude these cases to speed up the computation.

\subsection*{Second-order norms}

Although we have focused on the third-order norms, second-order norms are included in the third-order CESS's as a subset.
Under a second-order norm, a new reputation is assigned to a donor, and the prescribed action is independent of the donor's reputation, and
the assignment and action rules are functions of the recipient's reputation and the conducted action.
These norms are thus simpler than third-order ones.

The full list of the second-order CESS's is shown in Table~\ref{tab:second_order}.
We represent the norms by using the same notation as those in the main text, but with the first character unspecified (denoted as $\_$)
because the second-order norms are independent of the donor's reputation.

As shown in the table, there are $18$, $9$, and $6$ second-order norms in C1-P1-R1, C1-P2-R1, and C3-P1-R1 classes, respectively.
Most of these are relevant to norms with the binary reputation.
The only two second-order norms in the leading eight are Simple Standing (SS) and Stern Judging (SJ), which are denoted as ($\_B{:}DG{:}{\ast}, \_G{:}CG{:}B$).
As seen in Table~\ref{tab:second_order}, a large fraction of the norms, those denoted by $\spadesuit$ or $\clubsuit$,
are equivalent to SS or SJ when two of the reputations are merged into one.

\begin{table}[ht]
  \centering
  \caption{
  List of the second-order norms that are included in the CESS's.
  The asterisk $\ast$ represents a wildcard, and the square bracket $[BN]$ represents either $B$ or $N$.
  Those denoted by $\spadesuit$ are equivalent to SS or SJ
  when $B$ and $N$ are merged into a single reputation.
  Those denoted by $\clubsuit$ are equivalent to SS or SJ
  when $N$ and $G$ are merged into a single reputation.
  }
  \label{tab:second_order}
  \begin{tabular}{|ccc|c|c|} \hline
  \multicolumn{3}{|c|}{Prescriptions} & Type & Remark \\ \hline
  $\_B{:}DG{:}{\ast}$ & $\_N{:}DG{:}{\ast}$ & $\_G{:}CG{:}B$ & C1-P1-R1 & $\spadesuit$ \\ \hline
  $\_B{:}DG{:}{\ast}$ & $\_N{:}CG{:}B$ & $\_G{:}CG{:}B$      & C1-P1-R1 & $\clubsuit$ \\ \hline
  $\_B{:}DG{:}{\ast}$ & $\_N{:}DB{:}[BN]$ & $\_G{:}CG{:}B$   & C1-P1-R1 & \\ \hline
  $\_B{:}DN{:}[BN]$ & $\_N{:}DG{:}{\ast}$ & $\_G{:}CG{:}B$   & C1-P2-R1 & \\ \hline
  $\_B{:}DN{:}{\ast} $ & $\_N{:}CG{:}B$ & $\_G{:}CG{:}B$     & C1-P2-R1 & $\clubsuit$ \\ \hline
  $\_B{:}DG{:}{\ast} $ & $\_N{:}CG{:}B$ & $\_G{:}CN{:}B$     & C3-P1-R1 & $\clubsuit$ \\ \hline
  $\_B{:}DG{:}{\ast} $ & $\_N{:}CN{:}B$ & $\_G{:}CG{:}B$     & C3-P1-R1 & $\clubsuit$ \\ \hline
  \end{tabular}
\end{table}

\subsection*{Dynamics of C3 norms}
There are different kinds of dynamics of $h_N$ within C3 norms.
The fraction of $B$-players is a small quantity of $O(\mu)$ because one can easily escape from $B$-reputation by meeting $G$- or $N$-players.
If we merge $G$ and $N$ into a single reputation, the merged reputation
corresponds to $G$ for the leading eight. A radical example is a mechanism found
in $2,139$ norms, for which $h_G$ and $h_N$ are almost non-interacting: Just as
$G$-players preserve their reputations through $GG{:}CG{:}B$, the same is true
for $N$-players with $NN{:}CN{:}B$, and the interaction between $G$- and
$N$-players cannot change their numbers because the reputations are either
preserved or swapped. Their interaction is basically mediated by $B$-players,
originating from error $\sim O(\mu)$. Therefore, for each of these $2,139$
norms, the dynamics towards a fixed point becomes frozen as $\mu \to 0$.

For the rest, the coupling between $h_G$ and $h_N$ is more explicit, and the convergence rate is a finite constant independent of $\mu$. Let us give two representative examples:
A common pattern among $2,754$ norms in C3 is an oscillation between $G$ and $N$ due to $GG{:}CN{:}B$ and $NN{:}CG{:}B$. In the absence of error, two $G$-players will change their reputations to $N$, and vice versa. The master equation for $h_N$ is thus written as
\begin{equation}
\frac{d}{dt}h_N \propto h_G^2 - h_N^2 + O(\mu) = 1-2h_N + O(\mu),
\label{eq:osc}
\end{equation}
where we have plugged $h_G = 1-h_N$, considering $h_B \ll 1$. The above equation clearly shows $h_N^\ast = 1/2$ in the limit of small $\mu$.

For other $3,003$ of the C3 norms, the interaction between $G$ and $N$ is more delicate. In addition to the above prescriptions for Eq.~\eqref{eq:osc}, they also have $GN{:}CG{:}B$ and $NG{:}CG{:}B$ in common. Therefore, when an $N$-player meets a $G$-player, both will earn $G$-reputation by choosing $C$. As a whole, these features lead to the following master equation:
\begin{equation}
\frac{d}{dt}h_N \propto h_G^2 - h_N^2 - h_G h_N + O(\mu),
\end{equation}
whose stationary value is obtained as $\lim_{\mu \to 0} h_N^\ast = \varphi^{-2} = (3-\sqrt{5})/2 \approx 0.38$, where $\varphi \equiv (\sqrt{5}+1)/2$ is the golden ratio. The dynamical behaviours of the other C3 norms can be explained in similar ways. %such as
%\begin{equation}
%\frac{d}{dt}h_N \propto h_G^2 - h_G h_N + O(\mu),
%\end{equation}
%or
%\begin{equation}
%\frac{d}{dt}h_N \propto - h_N^2 + h_G h_N + O(\mu).
%\end{equation}

Table~\ref{tab:limit} summarizes our dynamical characterization of the three classes.

\begin{table}[t]
    \centering
    \begin{tabular}{|c|c|c|}\hline
    Types & Stationary values & Time dependence \\
         &  $h_N(\mu, t\to\infty)$ & $h_N(\mu \to 0, t)$ \\\hline
    C1     & $O(\mu)$ & $\exp(-t/\tau_N)$ \\\hline
    C2     & $O(\mu^{1/2})$ &  $t^{-1}$ \\\hline
    \multirow{3}{*}{C3}   & $1/5+O(\mu)$ & $\exp(-\mu t/\tau_N)$ \\
                          & $1/2 + O(\mu)$ & $\exp(- t/\tau_N)$ \\
                          & $\varphi^{-2} + O(\mu)$ & $\exp(- t/\tau_N)$ \\\hline
    \end{tabular}
    \caption{Summary of stationary values of $h_N^{\ast}$ and the asymptotic
    time evolution near stationarity.
    In every case, the common findings are $h_B(\mu, t\to \infty) = O(\mu)$ and $h_B (\mu\to 0, t) \sim \exp(-t/\tau_B)$. If necessary, a certain time scale of $O(1)$ is denoted by $\tau_N$ or $\tau_B$, which may differ norm by norm, and $\varphi \equiv (\sqrt{5}+1)/2$ is the golden ratio.}
    \label{tab:limit}
\end{table}

\section*{Acknowledgements}
Y.M. acknowledges support from Japan Society for the Promotion of Science (JSPS) (JSPS KAKENHI; Grant no. 18H03621 and Grant no. 21K03362).
S.K.B. acknowledges support by Basic Science Research Program through the National Research Foundation of Korea (NRF) funded by the Ministry of Education (NRF-2020R1I1A2071670).
Part of the results is obtained by using the Fugaku computer at RIKEN Center for Computational Science (Proposal number ra000002).
We appreciate the APCTP for its hospitality during the completion of this work.
This work is partially supported by RIKEN R-CCS International Student Internship.

\section*{Author contributions statement}
Y.M. designed the research, carried out
the computation, and analysed the results.
M.K. curated data and conducted formal analysis.
S.K.B. verified the method. Y.M.
and S.K.B. wrote and reviewed the manuscript.

\section*{Additional information}

The authors declare no competing financial interests.

\section*{Data Availability}
The source code for this
study is available at
{\url{https://github.com/yohm/sim_game_ternary_reputation}.

\end{document}